01.12.05

# A Study of the Operation of Especially Designed Photosensitive Gaseous Detectors at Cryogenic Temperatures

L. Periale[a], V. Peskov[b], C. Iacobaeus[c], B. Lund-Jensen[d], P. Pavlopoulos[b], P. Picchi[a], F. Pietropaolo[a]
[a]CERN, Geneva, Switzerland, CH-1211
[b]Pole University Leonard de Vinci, Paris, La Defense Cedex, France, 92916
[c]Karolinska Institute, Stockholm, Sweden, S-17176
[d]Royal Institute of Technology, Stockholm, Sweden, S-10691

## Abstract

In some experiments and applications there is need for large-area photosensitive detectors to operate at cryogenic temperatures. Nowadays, vacuum PMs are usually used for this purpose. We have developed special designs of planar photosensitive gaseous detectors able to operate at cryogenic temperatures. Such detectors are much cheaper PMs and are almost insensitive to magnetic fields. Results of systematic measurements of their quantum efficiencies, the maximum achievable gains and long-term stabilities will be presented.

The successful operation of these detectors open realistic possibilities in replacing PMs by photosensitive gaseous detectors in some applications dealing with cryogenic liquids; for example in experiments using noble liquid TPCs or noble liquid scintillating calorimeters.

## 1. Introduction

Noble liquids such as LAr, LKr and LXe are unique detecting medias: 1) their stopping power is high enough for many applications, 2) they are excellent scintillators, emitting in the VUV region of spectra, 3) primary electrons created inside the liquid could easily be drifted and collected on an electrode structure providing one with a charge signal, 4) if necessary, primary electrons could even be extracted from the liquids to the gas phase (vapours above the liquid layer) and collected on electrodes placed there. These properties make them attractive for several applications, for example noble liquids scintillating calorimeters [1], cryogenic TPCs [2], cryogenic PETs [3]. Nowadays, expensive PMs are used for the detection of the scintillation light in these devices [4]. There have also been some attempts to use solid –state detectors [5].

We have recently demonstrated that costly PMs and solid-state devices could be replaced by gaseous photosensitive detectors: detectors with a window (which could be immersed inside the noble liquids) or windowless, able to operate in pure noble gases or vapors above the noble liquids [6]. The advantages of these detectors are: a large sensitive area, the possibility to choosing construction materials with low radioactivity levels and the practical insensitivity to the magnetic fields. Due to these properties gaseous detectors are now considered as an option for XENON [7] and ZEPLIN [8] WIMPs detectors and cryogenic PETs [9].

The aim of this paper to develop prototypes of photosensitive gaseous detectors oriented on real experiments and perform the systematic studies of these devices.

## 2. Detector's Designs and Experimental Set Up

Two types of planar photosensitive detectors were constructed and tested in the frame of this work: the wire- type and the hole- type detectors. The schematic drawing of one of our wire detector (WD) mostly used in this work show in Fig. 1 (the description of other wire-type detectors can found in [6]).

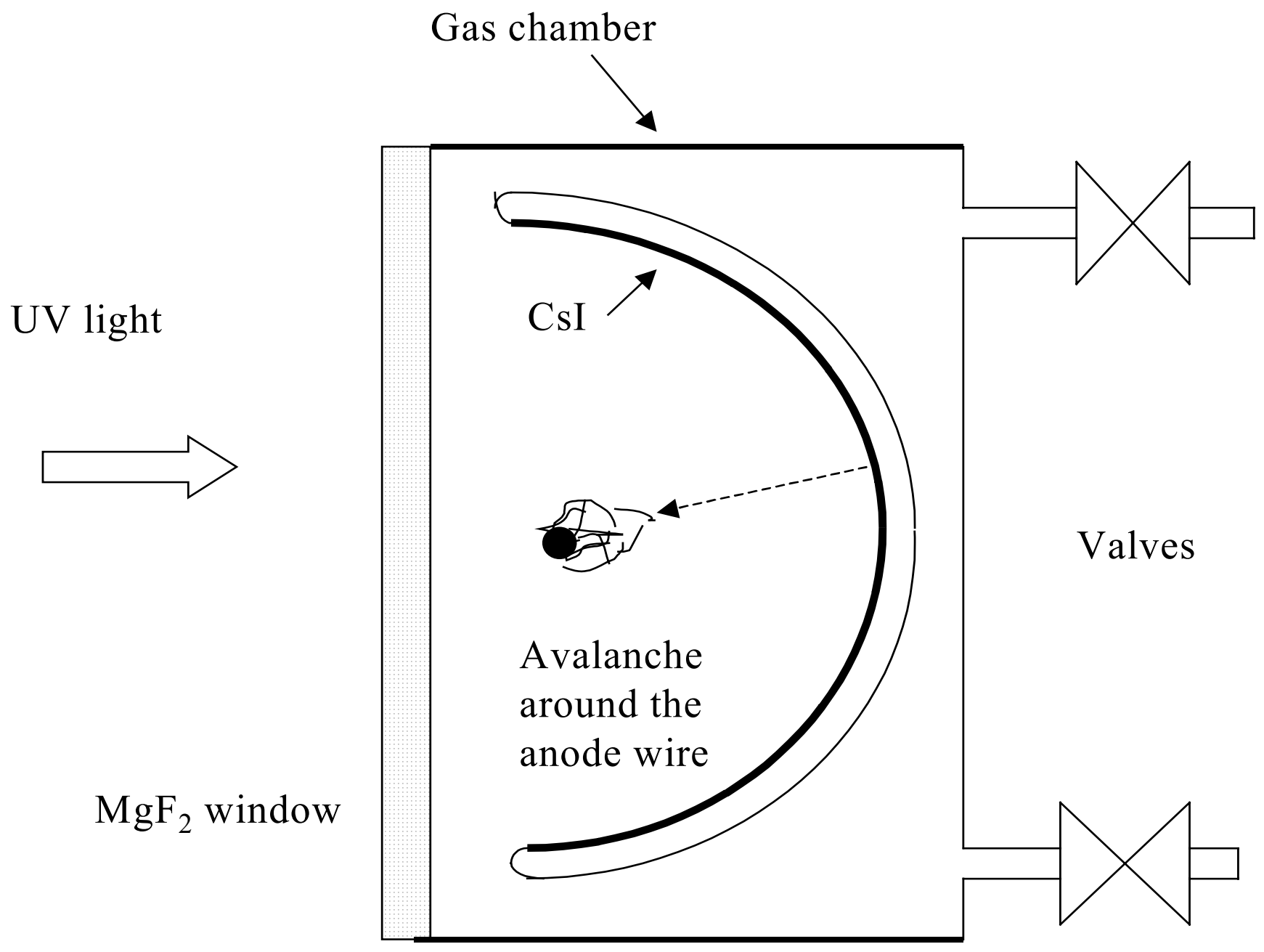

Fig. 1. Schematic drawing of the wire detector installed inside the gas chamber.

Its cathode was made from two parts: a stainless steal (ss) semi cylinder having a diameter of 2, 5 cm and a length of 5 cm was combined with a metalized $MgF_2$ window. Inside this structure a golden-coated tungsten anode wire was installed. The diameters of the anode wires tested were of 50 and 25 μm. The inner part of the ss semi cylinder was coated by a CsI layer 0,4mm thick. The distance between the semi cylinder and the $MgF_2$ window was 1 mm. This structure. The cathode parts were grounded; the high voltage was applied to the anode wire. This entire structure was placed inside a compact planar gas chamber (with a diameter of 10 cm) and filled either with Ar+10%$CH_4$ or He+10%$H_2$ at a total pressure p=1.

The hole- types detectors used in most these studies were either capillary plates (CPs) obtained from Hamamatsu or "home made CPs" (HMCPs) manufactured by us. The Hamamatsu CPs had a thickness of 0,8 mm and a diameter of 20 mm, whilst the diameter of the capillaries was 100 μm. Most of the measurements in this work were done with double CPs operating in cascade mode (see Fig. 2). The distance between the CPs (we named them "top" and bottom") was 2 mm. The cathode of the top CP was coated with a CsI layer 0,25 μm in thickness. The anode of the bottom CP was in direct contact with the readout plate.

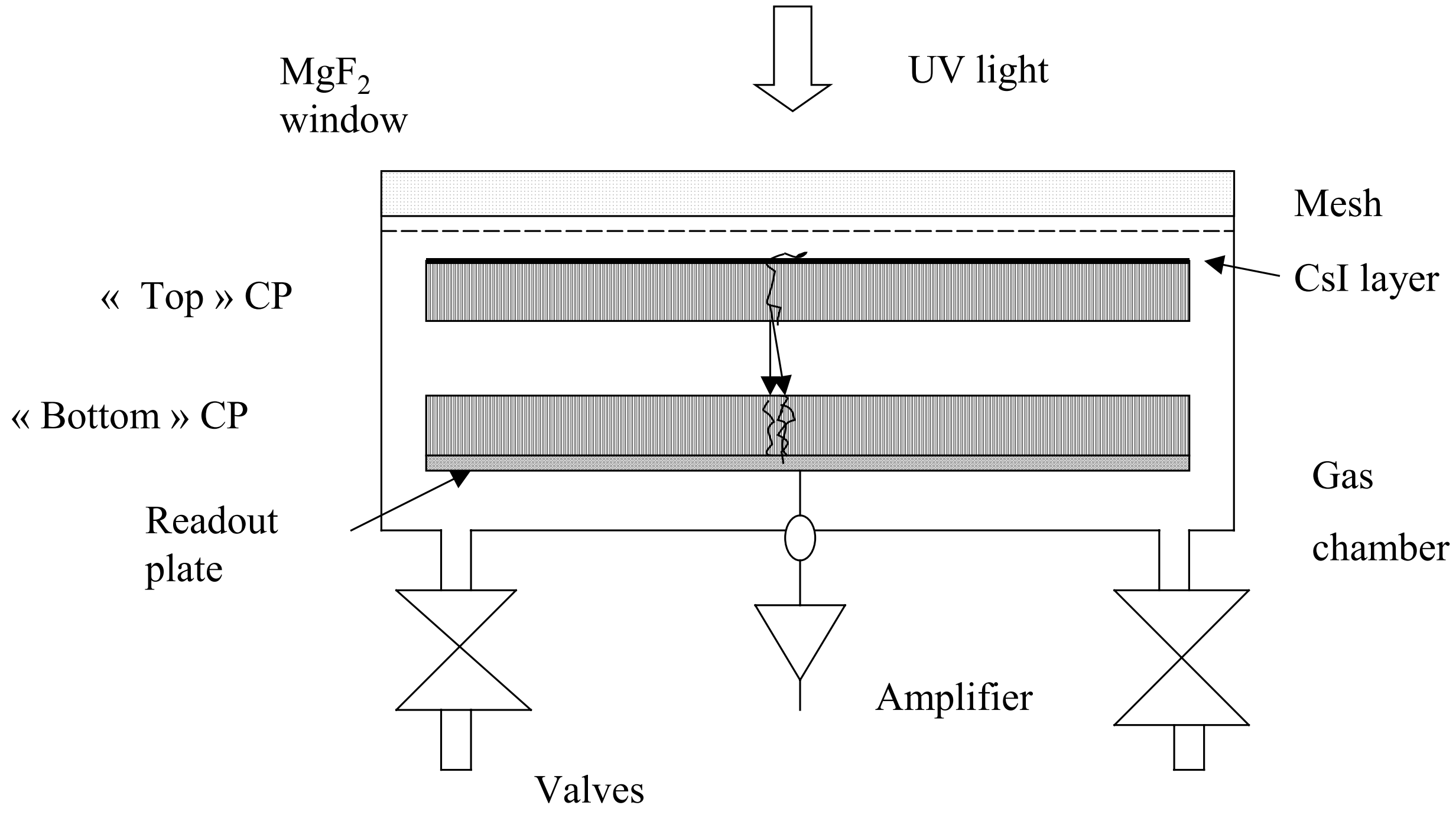

Fig. 2. Schematic drawing of the hole-type detector installed inside the gas chamber.

In some particular measurements, for example under conditions with a high risk of sparking, (HMCPs) were used. They were made of G-10, had thickness of 1mm, a diameter of 20mm whilst the diameters of the holes were of 0,3 mm. The advantages of these detectors are the very low price and the possibility of achieving gains higher then with CPs. We also used these detectors in a cascaded mode. The cathode of the top HMCP was coated by a CsI layer 0,4 μm in thickness.

The cascaded hole-type detectors were installed inside a small gas chamber filled either with He+10%$H_2$ (in the case of CPs) or with pure noble gases: Xe or Ar (as in the case of HMCPs) at a p=1.

In all experiments the gas chambers were attached to other chambers which we named a "scintillation chamber"(see Fig.1 in [6a]). This chamber was filled with on of the noble gases: Xe, Kr or Ar and contained a radioactive source ($^{241}$Am, $^{106}$Ru, $^{109}$Cd or $^{55}$Fe). These sources produced scintillation lights recorded by the photosensitive gaseous detectors. In some experiments the window separating the gas and the scintillation chamber was removed, so that the detector in the gas chamber was filled by the same noble gas as the scintillation chamber.

These two chambers coupled to each other were installed inside the cryostat allowing controllable cooling to be made from room temperature until $LN_2$ temperature (see [6c] for more details). While being located in the cryostat, the detectors could be run in two modes: flushed by a gas so that the pressure was kept at 1 atm, ("flushed detector") or filled by a gas to a p=1 atm at room temperature, sealed and then cooled so that the density remained constant during cooling and all measurements taken ("sealed detector"). In some measurements our gas chambers were directly immersed inside the $LN_2$, LAr or cooled alcohol.

Besides the scintillation light produced by the radioactive sources, in some measurements we also used external UV sources: a pulsed $H_2$ lamp (a few ns pulse durations) and continuous Hg lamp. The pulse lamp was very convenient in measuring the detector's gain and feedbacks [10]. The Hg lamp was used for measurements with single photoelectrons.

The absolute quantum efficiencies (QE) of our detectors were measured at room temperatures using a monochromator combined with a continuous Hamamatsu $H_2$ lamp. The intensity of the spectral resolved light beam from the $H_2$ lamp was measured by an ionization chamber filled with TMAE vapors in which the QE is well known. After these measurements, the relative changes in the QE were continuously monitored (this time without the monochromator) during all measurements (cooling, warm up) by measuring a photocurrent or a counting rate produce by the Hg lamp.

## 3. Results

The gain vs. voltage for the WD filled with $Ar+CH_4$ is presented in figure Fig 3. One can see that within this mixture the maximum achievable gain was very high: ${}^{WD}A_m > 10^6$. In spite of the fact that this detector could operate close to the limited Geiger mode, at gains of ${}^{WD}A_f > 5 \times 10^4$ photo feedback appeared. Note that for applications dealing with rare events (such as WIMP search or ICARUS) the feedback in photo-detectors was not a serious problem since the afterpulses appeared with a delay and could easily be excluded from the data processing. In these applications it is much more important to reach a high gain allowing one to achieve 100% efficiency for the detection of the photoelectrons.

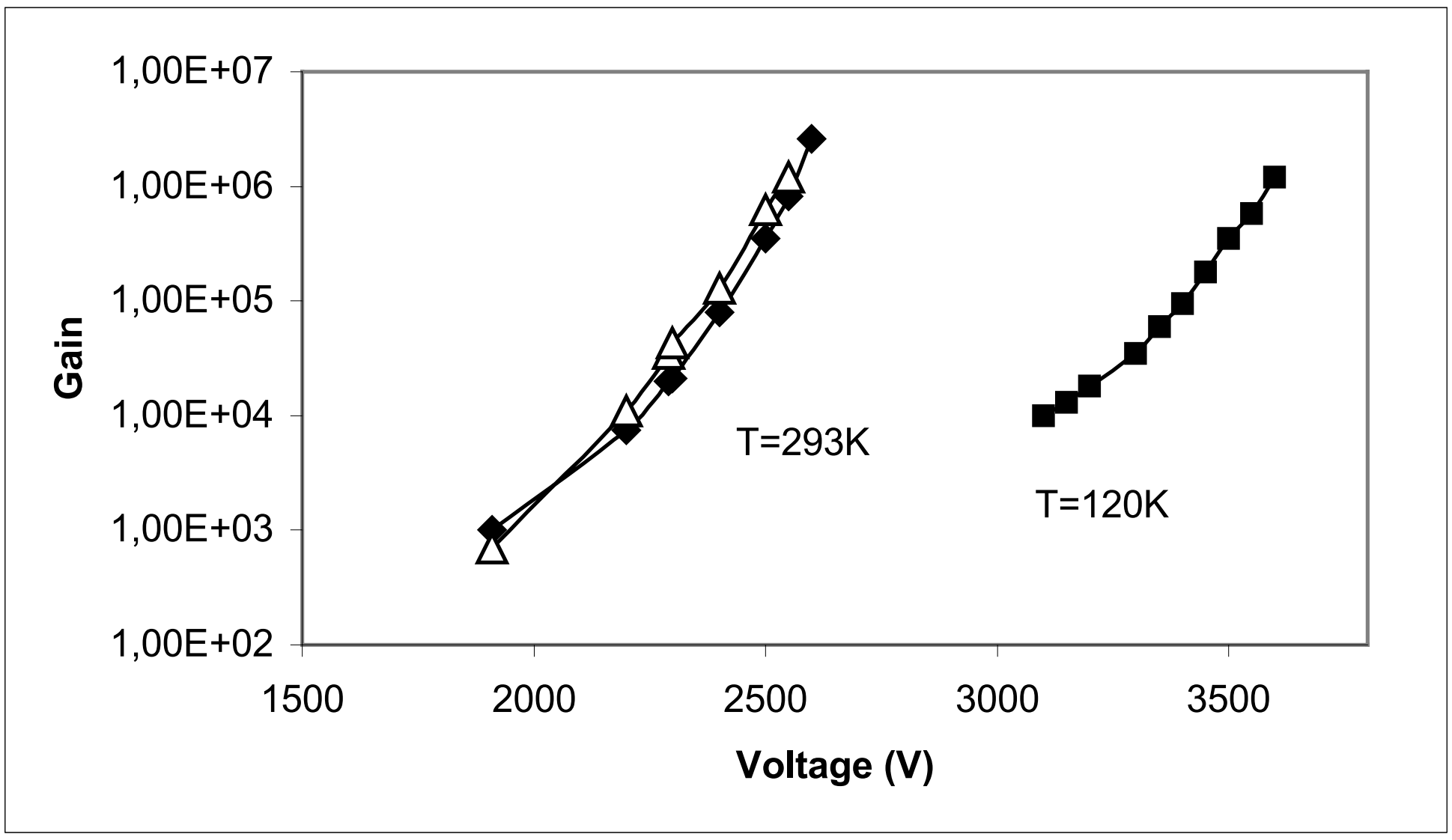

Fig. 3.Gain vs. voltage for a wire- type detector filled with $Ar+10\%CH_4$ gas mixture in a flushed mode (solid symbols) at temperatures of T= 293 K and 120 K, and in a sealed mode (open triangles) at T=120 K.

In $He+H_2$ mixture the maximum achievable gain ${}^{WD}A_m < 10^6$. The peculiarity of this mixture was that at low temperatures the flushed detector with the anode wire of 50 μm in diameter could transit from an avalanche to a self-quenched streamer mode-see Fig. 4. This effect is well known and was connected to the large diameter of the anode wire used and with the elevated density of the gas when it cooled [11]. In the case of a thin wire or in the case of the sealed detector, it operated in avalanche mode at any temperatures tested in this work. The feedback in the sealed appeared at ${}^{WD}A_f > 10^4$.

Thus one can conclude that in the case of the WD in both gas mixtures and at any given conditions ${}^{WD}A_m > {}^{WD}A_f$. As expected, in both mixtures the gain of sealed detector practically did not changed with the temperature.

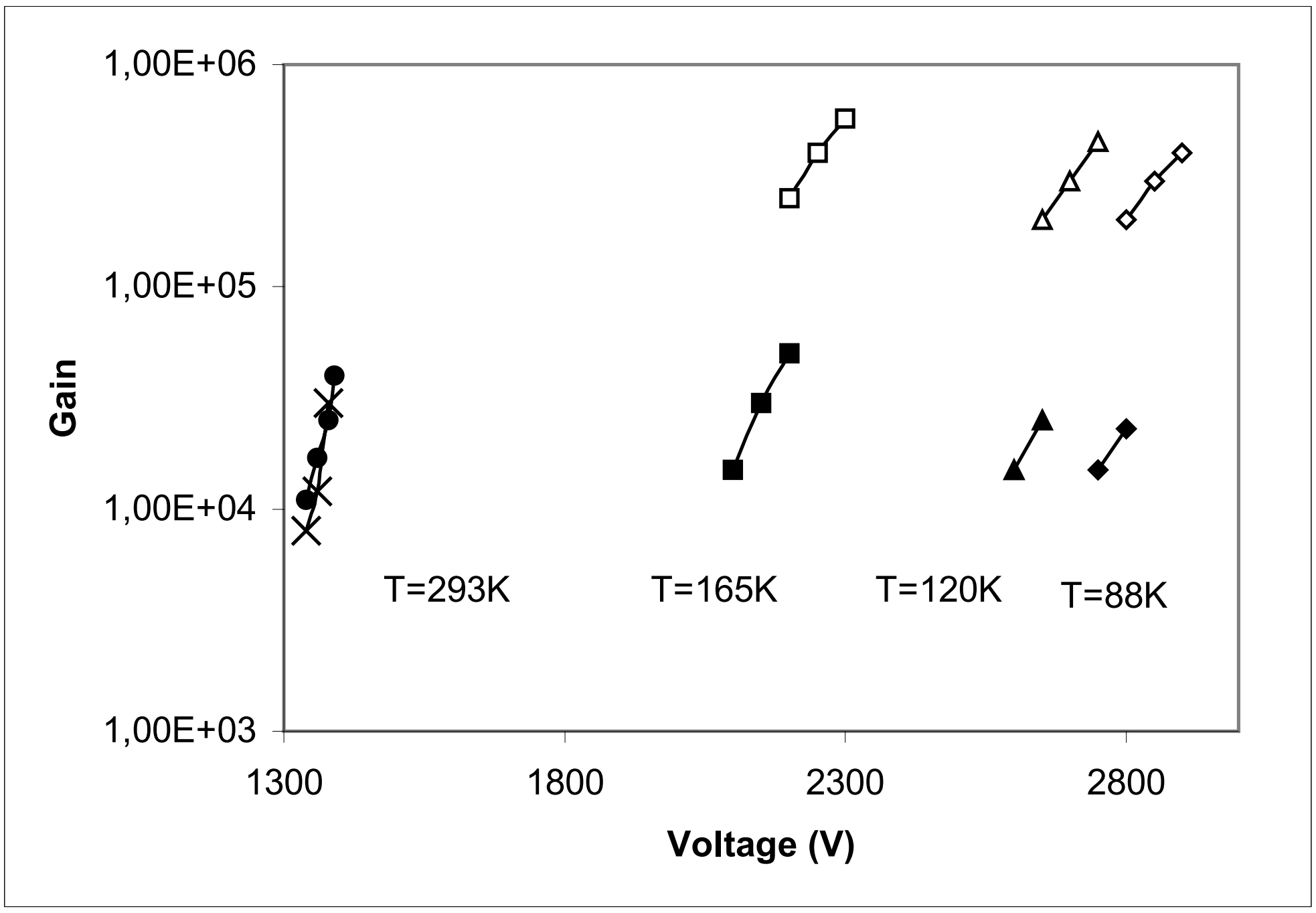

Fig. 4.Gain vs. voltage for the wire- type detector filled with a He+10%$H_2$ gas mixture in a flushed mode (open and filled symbols) at temperatures of 293 K, 165 K, 120 K, 88 K and in a sealed mode (crosses) at 88 K.

Results of the measurements the QE for WD, CPs and HMCPs are shown in Fig. 5. One can see that in He+$H_2$ gas mixture the QE is almost ten times lower than in the case of the Ar+$CH_4$ mixture. This is due to the strong backdiffusion of the photoelectrons in this mixture [6a].

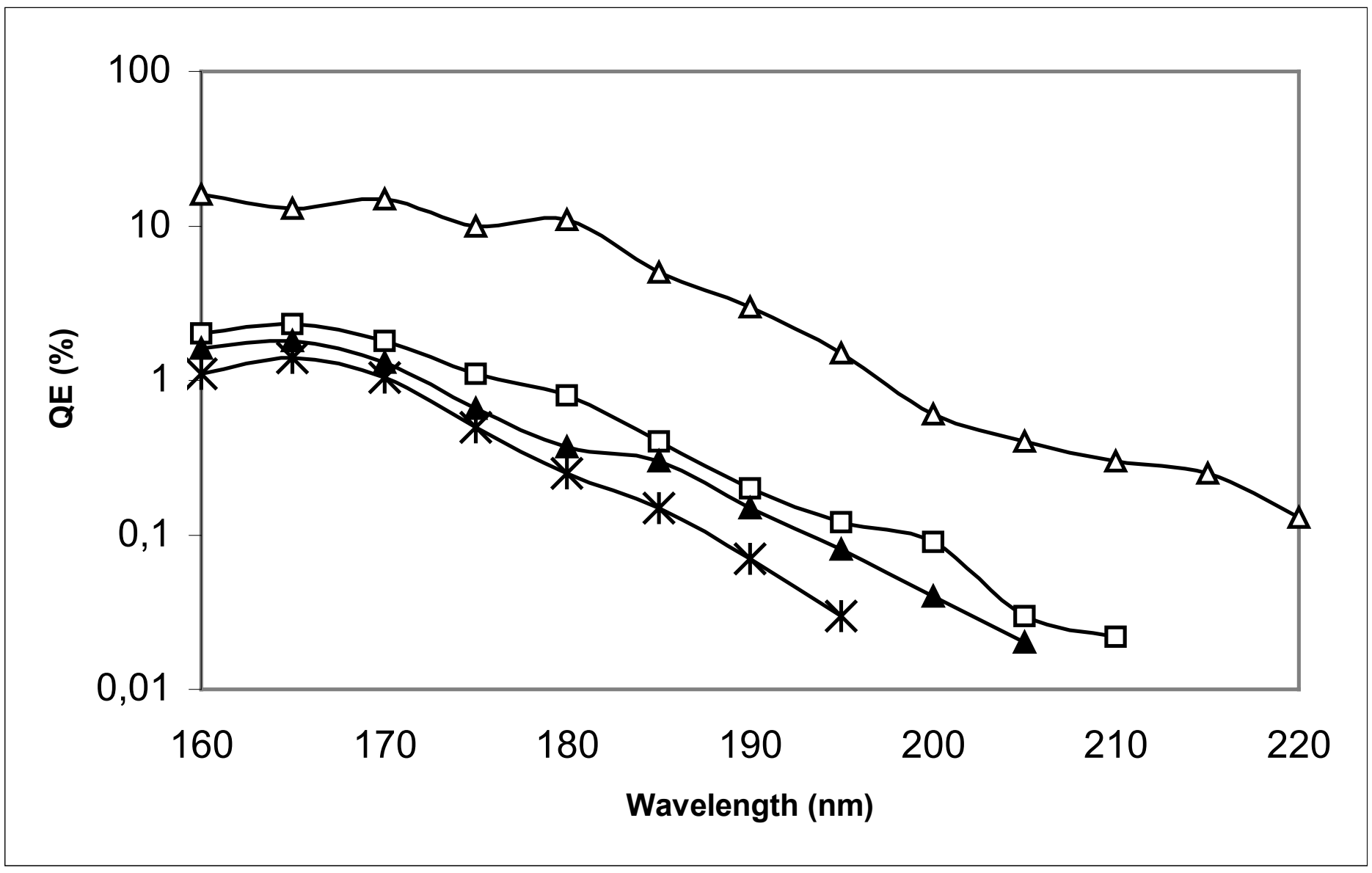

Fig. 5. The QE measured at room temperature for wire -and hole- type detectors: open triangles-WD filled with Ar+10%$CH_4$, open squares-WD filled with He+10%$H_2$, filled triangles –the QE of the CP in He+10%$H_2$, stars-the QE of HMCPs in Xe. In all measurements the total pressure of the gas mixture was 1 atm.

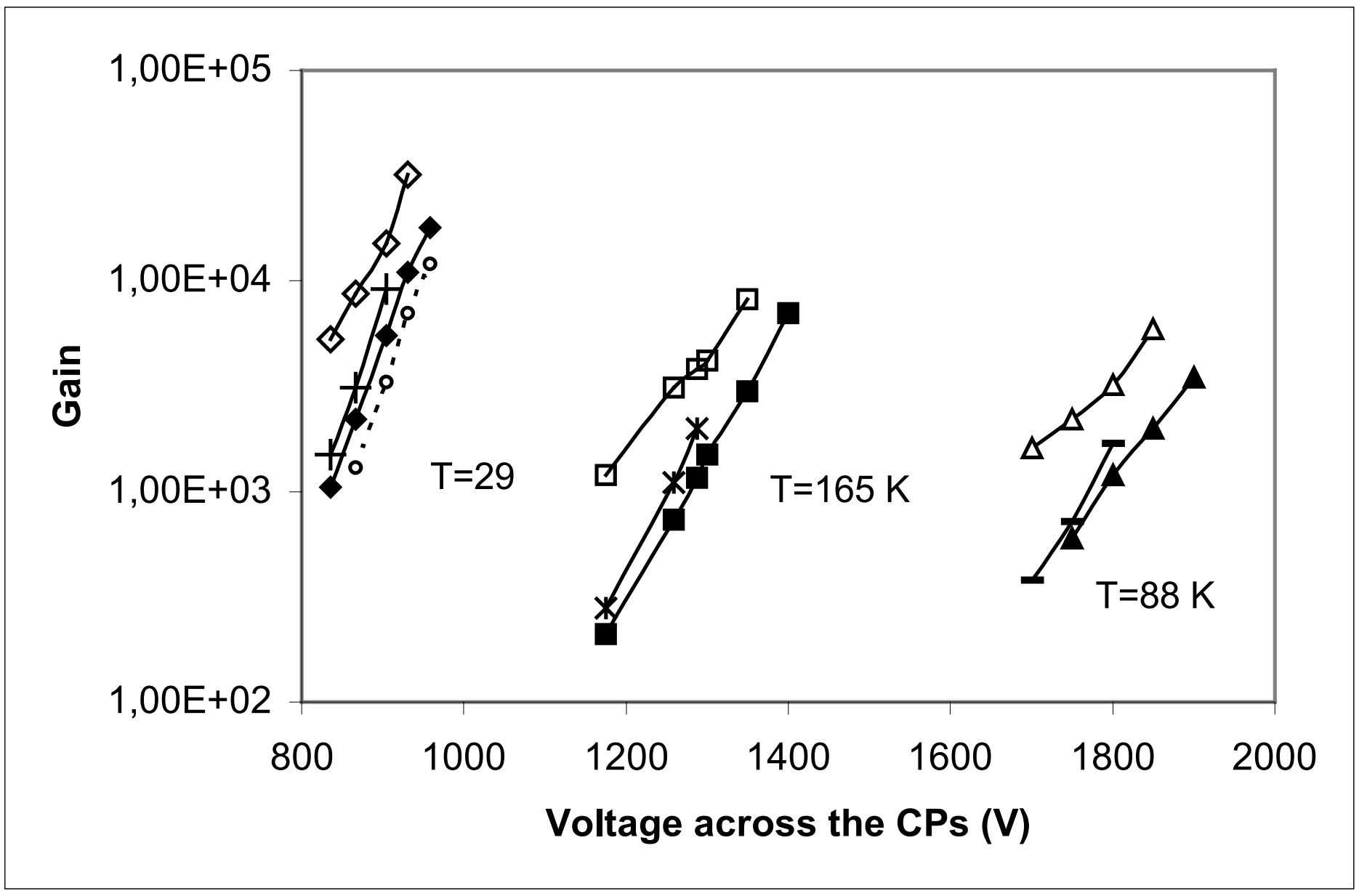

Fig. 6.Gain vs. voltage for CPs filled with He+10%$H_2$ gas mixture in flushed and sealed mode at various temperatures. Flushed detectors: solid symbols-single CPs, open symbols-two CPs operating in cascade mode. Crosses, stars and bars- double CPs with CsI photocathode. Dash line- a sealed single CP at 88 K

The gain vs. voltage for the CPs is presented in Fig. 6. One can see that the maximum achievable gain of the cascaded CPs is much lower than for WDs: $^{CP}A_m < 10^5$, however CPs operated without any feedbacks up to gains of up to those close to breakdowns: $^{CP}A_f \sim {}^{CP}A_m$. Thus CPs are preferred in applications when one has to avoid afterpulses caused by feedbacks.

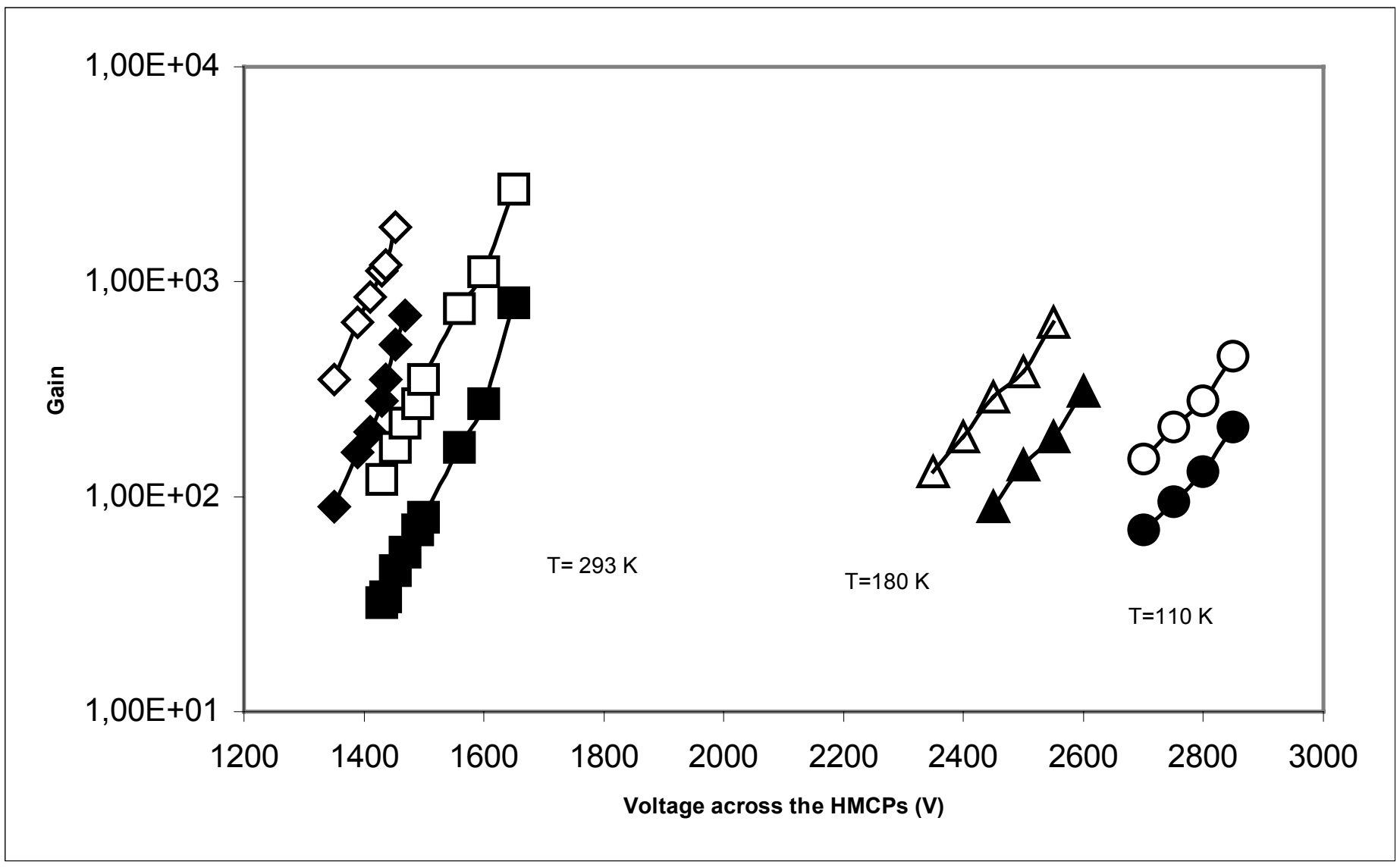

Fig. 7. Gain vs. voltage for single (solid symbols) and cascaded (open symbols) HMCPs in Xe (squares and triangles) or Ar (rhombus and circles) in flushed mode.

As was mentioned above, in some measurements the window was removed in order for the hole-type detectors to operate in the same noble gases as the scintillation chambers. In these measurements one has to avoid damages of the expensive CPs by possible sparks so HMCPs have to be used. As an example Fig. 7 shows the gain vs. voltage for HMCPs operating in Xe and Ar. One can see that with double HMCPs operating in a cascade mode one can reach gains of ${}^{HMCP}A_m \sim 10^3$. In this case we also observed that ${}^{HMCP}A_m \sim {}^{HMCP}A_f$.

All of our detectors were cooled and warmed up to room temperatures several times. The QE of the detectors was continuously monitored during all of these manipulations for a total time of three months. During this time interval the total time when the detectors were cooled was 50-80 hours. The results of the QE monitoring could be summarized as follows. For sealed detectors the changes in the QE was not more than a few %. For flushed detectors, the measured QE degraded with temperature to about 10-20%. This was simply due to the increase of the photoelectrons back diffusion with the gas density. Indeed, the same effect was observed when these detectors were pressurized at room temperature.

**4. Conclusions**

For the detection of the scintillation light from the LXe one can use WDs filled with $Ar+CH_4$ gas mixture which allow one to achieve a high gain and a high QE. For the temperatures below 120K, to avoid the liquidation of the gas inside the WD, one has to use $He+H_2$, and the price to pay for this is the lost of the QE due to the strong back diffusion effect. In both gas mixtures WDs can operate at higher gains than CPs; however, CP operated without any feedback afterpulses. These results could be summaries as follows: ${}^{WD}A_f < {}^{CP}A_m \sim {}^{CP}A_f < {}^{WD}A_m$.

In pure noble gases HMCPs worked quite stabile. In spite the fact that their QE was not very high, one can compensate this by using large- area detectors.

Results obtained in this work indicate that photosensitive gaseous detectors (with windows and without) offer a cheap and simple alternative approach to PMs and solid -state detectors. This may open their applications to large-scale detectors such as cryogenic TPCs, PETs and noble liquid calorimeters.

**References:**

[1] M. Chen et al., Nucl. Instrum. and Meth. in Phys. Res. A 327 (1993) 187.
[2] C. Montanari et al., Nucl. Instrum. and Meth. in Phys. Res. A 518 (2004) 216.
[3] V. Chepel et al., Nucl. Instrum. and Meth. in Phys. Res. A 392 (1997) 427.
[4] Talks at the Second Workshop on Large TPC for Low Energy Rare Events Detection, Paris, France, Dec. 2004, http://www.unine.ch/phys/tpc.html.
[5] A. Braem et al., Nucl. Instrum. and Meth. in Phys. Res. A 320 (1992) 228; K. Ni et al., Preprint Physics/0502071, Febr. 2005.
[6] (a) L. Periale et al., Nucl. Instrum. and Meth. in Phys. Res. A 535 (2004) 517;
(b) L. Periale et al., Preprint Physics/0403087, Febr. 2004 ;
(c) L. Periale et al., Preprint Physics/0410280, Nov. 2004.
[7] E. Aprile, “XENON: a Liquid Xe Experiment for Dark Matter” Proposal # 0201740 for Nat. Science Foundation , Columbia University, NY, USA, September 26, 2001.
[8] http://www.shef.ac.uk/physics/research/pppa/research/ukdmc/ZEPLIN.html.
[9] C. Grignon et al., “Simulation of a High Performance γ- Camera Concept for PAT based on LXe and Gaseous Photomultipliers”- report at the 15th IEEE International Conference on

Dielectric Liquids, Coimbra, Portugal, June 2005, will be available on-line at the IEEE Xplore.
[10] C. Iacobaeus et al., Nucl. Instrum. and Meth. in Phys. Res. A 525 (2004) 42.
[10] P. Fonte et al., ICFA Instrum. Bull. Vol. 15, SLAC-PUB-77 and SLAC-JOURNAL-ICFA-1518 Fall. Issue 97.